\documentclass[%
 aip,
 amsmath,amssymb,
 reprint,%
]{revtex4-2}

\usepackage{amsmath}
\usepackage{graphicx}
\usepackage{dcolumn}
\usepackage[dvipsnames]{xcolor}
\usepackage{bm}

\colorlet{blue}{black}
\colorlet{red}{black}
\colorlet{BurntOrange}{black}

\newcommand{\db}[1]{\textcolor{red}{#1}}
\newcommand{\lt}[1]{\textcolor{BurntOrange}{#1}}

\usepackage[utf8]{inputenc}
\usepackage[T1]{fontenc}
\usepackage{mathptmx}
\usepackage{etoolbox}

\makeatletter
\def\@email#1#2{%
 \endgroup
 \patchcmd{\titleblock@produce}
  {\frontmatter@RRAPformat}
  {\frontmatter@RRAPformat{\produce@RRAP{*#1\href{mailto:#2}{#2}}}\frontmatter@RRAPformat}
  {}{}
}%
\makeatother

\begin{document}

\title[The effects of solvent quality and core wetting on the circularization of star polymers]{The effects of solvent quality and core wetting on the circularization of star polymers}

\author{Davide Breoni}
\affiliation{Department of Physics
, Università di Trento, Via Sommarive 14, I-38123 Trento, Italy }
\affiliation{INFN-TIFPA, Trento Institute for Fundamental Physics and Applications, I-38123 Trento, Italy}
\author{Emanuele Locatelli}%
\affiliation{Department of Physics and Astronomy, University of Padova, Via Marzolo 8, I-35131 Padova, Italy}
\affiliation{INFN, Sezione di Padova, Via Marzolo 8, I-35131 Padova, Italy}
\author{Luca Tubiana$^*$}
\affiliation{Department of Physics
, Università di Trento, Via Sommarive 14, I-38123 Trento, Italy }
\affiliation{INFN-TIFPA, Trento Institute for Fundamental Physics and Applications, I-38123 Trento, Italy}
 \email{luca.tubiana@unitn.it}

\date{\today}

\begin{abstract}
We simulate the formation of cyclical arms in star polymers, focusing on the effects of solvent quality on their resulting linking complexity and gyration radius. We find that polymers circularized in bad solvent present a higher degree of linking among arms with respect to those circularized in good solvent. When both are transported to good solvent, this results in a smaller gyration radius of the former with respect to the latter. This effect is magnified when the polymers present a sufficiently small number of arms (or  functionality $f$): in this case, in bad solvent, all arms tend to clump together on one side of the central core, due to circularization, and can hence all interact with each other. Instead, when $f$ is large enough, the whole surface of the core is wetted by the arms, whose distribution becomes radially symmetric. This hinders interactions between faraway arms and reduces the probability of inter-arm linking. Interestingly, we find that both the critical $f_c$ at which the clump transition happens and the minimal arm length $n_c$ for which the transition appears depend on the core size: \db{the grafting density of the arms must be larger than a certain constant $\rho_g^c$, while their length must be sufficient to stretch for, at least, half of the core's circumference. }
\end{abstract}

\maketitle

\section{\label{sec:intro}Introduction.}
Star polymers --- macromolecules composed of several polymers, or arms, attached to a central core --- are of great interest in soft matter physics, chemistry, and material science. While, from the theoretical perspective, they can be described through scaling theories and allow to connect polymer physics and colloidal science~\cite{grest_star_1996,likos_effective_2001,vlassopoulos_colloidal_2004,ren_star_2016}, their well defined structure and the high level of control achievable during their synthesis allows their use in several chemical and technological applications, both in research and industry~\cite{vlassopoulos_tunable_2014}. The properties of star polymers depend heavily on the solvent quality in which they are immersed, as they undergo a coil-to-globule transition~\cite{huissmann_star_2009,verso_effect_2012,selli_impact_2019,a_farimani_coarse-graining_2025}, yielding interesting behaviors in flows~\cite{peng_reducing_2022}, and influencing inter-molecular interactions~\cite{zhang_spherical_2013,loverso_interaction_2012,dutta_quantification_2013,nikoubashman_coarse-graining_2015}, polymerization~\cite{frohlich_shielding_2010} and self-assembly~\cite{liao_self-assembly_2014}. This aspect raises the interesting question of whether one could quench some of the properties that star polymers have in bad solvent so that they are kept in good solvent as well. \lt{For example, one might hope to use this methodology to change the pair interaction between star polymers so that, in dense solutions, they form a different phase, while keeping the other properties characteristic of good solvent. In principle, this might be possible through topological constraints.}

Topological properties and constraints are defined by global features of the system that are maintained through continuous transformations, such as knotting and linking in cyclical polymers~\cite{tubiana_topology_2024}. They can be exploited to probe physical properties by analyzing the probability of occurrence of different topologies as a function of the conditions in which they are formed: for example, the probability of observing different knots has been used to estimate the salt-dependent excluded volume of DNA~\cite{rybenkov_probability_1993}, infer the genome organization inside a viral capsid~\cite{arsuaga_investigation_2002,arsuaga_dna_2005,marenduzzo_dna-dna_2009}, \lt{or characterize the behaviour of polymers in ultracentrifugation assays~\cite{cunha2025hierarchical}}. Further, the linking probability has been used, for example, to understand the structure and properties of the kinetoplast DNA~\cite{chen_topology_1995,he_single-molecule_2023,klotz_equilibrium_2020,ramakrishnan2025organisation}, \lt{of various gels and melts~\cite{palombo2025topological,stavno2025topology}} and the network  of bonds of tetrahedal liquids (such as water), a key ingredient for the understanding of their transitions~\cite{neophytou_topological_2022}. Once trapped in circularized macromolecules, knots and links cannot be removed without cutting the molecules open again, thus restricting their conformational space and affecting their physical properties. This aspect can be used to obtain materials with desirable \lt{equilibrium and dynamic properties properties~\cite{farimani2024effects,bracewell2025topology,sapsford2025topologically}, including olympic networks}~\cite{krajina_active_2018,klotz2024chirality,luengo2024shape,michieletto2025kinetoplast,zhang2025dna,speed2024assembling}, polycatenanes~\cite{wu_polyncatenanes_2017,ahmadian_dehaghani_effects_2020, liu2022polycatenanes,chiarantoni2022effect, tubiana_circular_2022,liu2023infinite}, and knotted molecules~\cite{fielden_molecular_2017,cardelli2018heteropolymer,chang_polymer_2005}. 

In principle, one can merge the two uses of topological properties, as probes or as constraints, by employing them to ``memorize'' some of the physical features, typical of a certain environment (solvent quality, degree and dimensionality of confinement...), making them available in other conditions. This is our approach here: we investigate the possibility of using topological constraints to create star polymers with ring arms, circularized in such a way to trap properties typical of bad solvent, like a smaller radius of gyration, and bring them to good solvent. We propose doing so through a click-chemistry approach. ``Click-chemistry'' is an umbrella term for any reaction in which  two molecular components, each equipped with appropriate functional groups, are joined~\cite{kolb_click_2001,meldal_polymer_2008,devaraj_introduction_2021}. This methodology has found already plenty of applications in biology~\cite{jewett_cu-free_2010, el-sagheer_click_2010}, medicine~\cite{kolb_growing_2003,hou_impact_2012}, material science~\cite{moses_growing_2007} and polymer physics~\cite{binder_click_2007,geng_click_2021}. 

Using an effective approach, in this study we consider linear stars with functionalized end groups (Fig.~\ref{fig:sticky}a) that ensure the formation of single bonds, leading to circularization (Fig.~\ref{fig:sticky}b). By means of this circularization, we can quench some of the properties of the conformation assumed by the star polymer at the time of the click-chemistry reaction.
We find that stars circularized in good solvent have little to no topological signature, regardless of their number of branches --- called functionality --- $f$. This is compatible with our previous study, Ref.~\cite{breoni_conformation_2024}, in which we observed that to have an effect on the physical properties of a star polymers, links should involve distant branches. Instead, in bad solvent conditions, the architecture plays a role: a transition is observed in the polymer conformations, with increasing functionality. Notably, the effects of a bad solvent circularization are visible even if the star is brought to good solvent conditions.
We elucidate the origin of this phenomenon and discuss it as a wetting transition: at low $f$ the arms wet the core only partially, giving rise to an asymmetric collapse. However, increasing $f$, the wetting becomes complete and the macromolecule becomes radially symmetric. We further characterize this transition varying the number of beads per arm $n$ and the core diameter $\sigma_c$, finding that the critical functionality $f_c$ depends on a constant critical grafting density $\rho_g^c$ (i.e. the number of arms per core surface unit) and is independent of $n$, for large enough values of $n$.

The paper is organized as follows: in Sec.~\ref{sec:methods} we describe the numerical model employed (Subsec.~\ref{ssec:simulations}), the initial configurations for the simulations (Subsec.~\ref{ssec:init}) and the quantities we considered in our characterization of the assembled cyclic stars: metric (Subsec.~\ref{ssec:asph}) and topological (Subsec.~\ref{ssec:linkn}). In the section dedicated to the main results, Sec.~\ref{sec:results}, we aim to characterize the conformational properties of a cyclical star dynamically formed via click-chemistry in three different settings: good solvent, bad solvent and bad-to-good solvent. In Subsec.~\ref{ssec:gyration} we study the gyration radius $R_g$ and asphericity $A$ of the stars, finding the hallmarks of a conformational transition in the bad solvent case. In Subsec.~\ref{ssec:rlink} we focus on the interplay between the transition and the topology of the star, in the form of the average linking per arm $\overline{|Lk|}$ and, in Subsec.~\ref{ssec:rasph}, we characterize the aforementioned transition, happening in bad solvent, as a function of the core diameter $\sigma_c$, the number $f$ and the length $n$ of the arms. Finally, in Sec.~\ref{sec:outro} we discuss the results and outlooks of the project.
\section{\label{sec:methods}Methods.}
\subsection{\label{ssec:simulations} Numerical model and simulation details.}
We perform coarse-grained simulations of 3D star polymers in a infinitely diluted bulk solution (that is, we consider a single macromolecule in a large simulation box) using LAMMPS~\cite{LAMMPS}. We consider star polymers with functionality $8 \leq f \leq 160$, each arm having $5\leq n \leq 50$ beads. The ending bead of each branch is reactive, that is, it is functionalized to enable bond formation with the corresponding bead of another arm. The core diameter is $\sigma_c=c\sigma$, where $\sigma$ is the diameter of all non-core beads and $6\leq c\leq 8$. Its mass is correspondingly set to $m_c=\left(\frac{\sigma_c}{\sigma}\right)^3m=c^3m$, where $m$ is the mass of a non-core bead. 
The polymers are modeled within the Kremer-Grest framework~\cite{grest_structure_1987}, where consecutive beads are connected via finite-extensible non-linear elastic (FENE) springs $U_s^{ab}(r)$, where $a$ and $b$ determine the type of the interacting particles (either core or non-core), shifting the potential of an amount $\Delta^{ab}$ to take into account the different radii:
\begin{equation}
\label{Us}
U_s^{ab}(r)=
\begin{cases}
-\frac{KR_0^2}{2}\text{ln} \left[1-\left(\frac{r-\Delta^{ab}}{R_0}\right)^2\right],&r\leq R_0+\Delta^{ab},\\
\infty, &r>R_0+\Delta^{ab},
\end{cases}
\end{equation}
where $r$ is the distance between beads, $\Delta^{ab}=\frac{1}{2} (\sigma_a+\sigma_b)-\sigma$ is the shift, $K=30\epsilon/\sigma^2$ is the stiffness of the spring, $\epsilon$ is the energy scale of the system and $R_0=1.5\sigma$ is the maximum bond length. The quantities $\sigma_a$ and $\sigma_b$ can assume the values of either $\sigma$ or $\sigma_c$. The arms are attached to a central core via anchoring monomers (or points), kept at a fixed position with respect to the core; we treat the collection of the anchoring points plus the central bead as a rigid body. For each bead $i$ with two neighbors (except the core), \db{a small bending energy $U_{b,i}$ is introduced to prevent unphysical configurations}:
\begin{equation}
\label{Ub}
U_{b,i}=
\kappa (1+\cos{\theta_i}),
\end{equation}
where $\theta_i$ is the angle formed by beads $i-1$, $i$ and $i+1$ and $\kappa=\epsilon$ is the bending energy. 
All beads interact with each other with a Lennard-Jones (LJ) potential $U_r^{ab}(r)$, where $a$ and $b$ determine the type of the interacting particles (either core or non-core), again shifted to accommodate different bead sizes:
\begin{equation}
\label{Ur}
U_r^{ab}(r)=
\begin{cases}
4\epsilon \left[\left(\frac{\sigma}{r-\Delta^{ab}}\right)^{12}-\left(\frac{\sigma}{r-\Delta^{ab}}\right)^6\right]+\epsilon_c, &r\leq r_{c}+\Delta^{ab},\\
0 ,&r> r_{c}+\Delta^{ab}.
\end{cases}
\end{equation} 
We change the quality of the solvent by varying the cutoff distance $r_{c}$ in eq.~\eqref{Ur}; we consider $r_{c}=\sqrt[6]{2}\sigma$ for good solvent and $r_{c}=3\sigma$ for bad solvent. The term $\epsilon_c$ depends on $r_c$ and is chosen such that $U_r^{ab}(r_c+\Delta^{ab})=0$.  It is important to remark that, in our model, the bad solvent condition implies the presence of an effective attraction not only between the beads but also between the beads and the core. 

In order to model binding between functionalized reactive beads, we use the potential proposed by Sciortino \cite{sciortino2017three}, that relies on a modified Stillinger-Weber three-body potential \cite{stillinger_computer_1985}.
The potential $U_{ck}$ is defined with the following set of equations:
\begin{align}
    U_{ck}&=\sum_i\sum_{j>i}u_2(r_{ij}) + \sum_i\sum_{j\neq i}\sum_{k>j}u_3(r_{ij},r_{ik});\\
    u_2(r_{ij})&= A\epsilon_b\left[B\left(\frac{\sigma_s}{r_{ij}}\right)^4-1\right]\text{e}^{\sigma_s /(r_{ij}-r_{cv})};\\
    u_3(r_{ij},r_{ik})&=0.9\epsilon_b u_v(r_{ij})u_v(r_{ik});\\
    u_v(r_{ij})&=\begin{cases}1 &\text{for } r_{ij}\leq r_{\text{min}}\\
    -u_2(r_{ij})/\epsilon_b &\text{for }  r_{\text{min}}<r_{ij} \leq r_{cv},
    \end{cases}
\end{align}
where $i,j,k$ refer to the indices of the ending monomers on different branches, $r_{\text{min}}$ is the distance at which $u_2(r_{ij})$ has a minimum, $\epsilon_b=15\epsilon$,  $\sigma_s=1.05\sigma$, $r_{cv}=1.68\sigma$, $B=0.41$ and $A=8.97$.  \db{The main feature of this potential is the three-body term $u_3$, which almost perfectly counterbalances the energy obtained when more than two particles bind. This renders triplet bonds unstable and allows bond rearrangement at virtually no energy cost (in all panels Fig~\ref{fig:sticky}c,d,e the binding energy remains constant). 
This potential has been employed in the modelisation of different polymeric systems, as it allows to avoid kinetic traps and to reach equilibrium in the limit of strong anisotropic attractive interactions~\cite{gnan2017silico,rovigatti2018self,heidenreich2020designer,paciolla2022validity,rovigatti_designing_2022}.
In our case, by ensuring the formation of single bonds between functionalized beads and, at the same time, allowing them to be exchanged during the run, the potential $U_{ck}$ fits the modelization of the binding process and lets us sample a very large number of topologies within a single run.  This effective approach embraces the spirit of click chemistry, which focuses on efficiency and simplicity\cite{devaraj2021introduction}.}

The equations of motion are integrated with a standard Velocity Verlet algorithm and the system is kept at fixed temperature by a Langevin thermostat, with temperature $T=\epsilon/k_B$ and damping coefficient $\gamma=\tau^{-1}$, where $\tau=\sigma \sqrt{m/\epsilon}$ is the unit of time of the system. The time-step is $\text{d}t=0.01\tau$. For each system, we simulate 8 independent realizations and perform long simulation runs of $10^6 \tau$, starting from well equilibrated non-reactive conformations, that is, where the three body potential is turned off (see below).   

\subsection{\label{ssec:init}Initial configurations and solvent properties.}
\begin{figure}[h]
\centering
\includegraphics[width=0.44\textwidth]{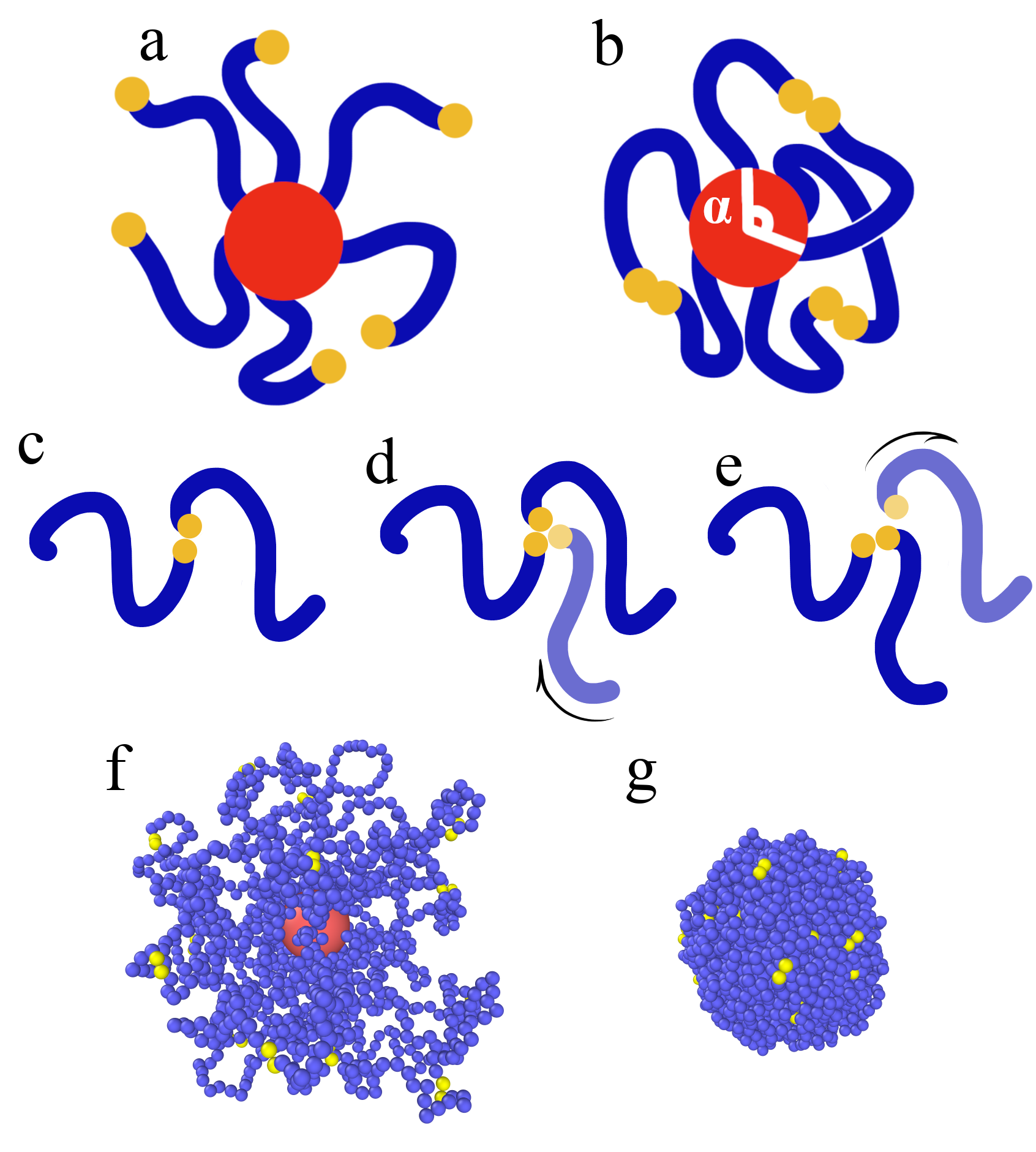}
\caption{(a-b) Sketch of circularization driven by click chemistry. The core is shown in red, the inert arm beads are blue, while the functionalized ones are yellow. The arms of a linear star (a) attach to each other via their functionalized ends, forming a circularized star (b). We define the angle formed by the grafting points of two attached arms as $\alpha$. (c-e) Sketch of bond exchange via three-body potential. Thanks to this potential, the bonding energy over the whole process (c,d,e) does not change, making a bond among three functionalized beads unlikely and at the same time allowing bond exchange. (f-g) Simulation snapshots of star polymers with core diameter $\sigma_c=6\sigma$ and number of beads per arm $n=40$ in different setups.  (f) Star polymer with number of arms $f=32$ in good solvent. (g) Star in bad solvent with $f=64$.} 
\label{fig:sticky}
\end{figure}

In order to cover the surface of the core in the most uniform way possible and to avoid conflicts between beads, we anchor the $f$ arms to the surface of the core following a spherical Fibonacci spiral (also known as {\it sunflower seed pattern}) \cite{gonzalez_measurement_2010}, as in our previous work~\cite{breoni_conformation_2024}. We start the simulations from an initial configuration in which the arms extend radially from the core, and equilibrate the system with non-reactive ending beads. Once thermal equilibrium is reached, we turn on the three-body potential, allowing the ends to circularize in a way that is independent of the initial configurations. 
We study their circularization in different conditions. First, we study what happens when stars are circularized in both good solvent (Fig.~\ref{fig:sticky}f) and bad solvent (Fig.~\ref{fig:sticky}g), allowing bonds to swap to dynamically sample assemblies with different topologies. We will denote in the following these two setups as {\it good} and {\it bad} respectively. Then, we also consider the setup in which we take the final configuration of stars circularized in bad solvent, fix the bonds permanently (quenching) and simulate them in good solvent; this will be denoted as {\it bad-to-good}.  
\subsection{\label{ssec:asph}Metric properties.}
We consider here two observables, specifically the {\it gyration radius} $R_g$ and the {\it asphericity} $A$. The gyration radius $R_g$ is defined as:
\begin{equation}
\label{rg}
    R_g \equiv \left(\frac{1}{N}\left\langle\sum_{i=1}^{N} (\textbf{r}_i-\textbf{r}_{0})^2\right \rangle\right)^{1/2} - \frac{\sigma_c}{2}, 
\end{equation}
where $N=f\cdot n$  is the total number of arm beads, $\textbf{r}_i$ is the position of $i$-th monomer, $\textbf{r}_0$ is the position of the core and $\langle\cdot\rangle$ is the ensemble average, estimated by computing over time along a single, long trajectory (see Sec.~\ref{sec:methods}) and over the ensemble of 8 independent stars per trajectory.\\
The asphericity is computed employing a modified moment of inertia tensor $T_{\alpha\beta}$, which we define as
\begin{eqnarray}
\label{Tab}
    T_{\alpha\beta} &\equiv& \frac{1}{N}\left\langle\sum_{i=1}^{N} (r_{\alpha,i}-r_{\alpha,0})(r_{\beta,i}-r_{\beta,0})\right \rangle,
\end{eqnarray}
where $\alpha$ and $\beta$ refer to the Cartesian axes. This modified tensor is calculated in the reference system of the core bead $\textbf{r}_0$, and not the center of mass, as we are interested in deviations from a spherical configuration centered in the core.\\
From $T_{\alpha\beta}$ we extract the semiaxes of the ellipsoid enveloping the star polymer by taking the square root of the three eigenvalues of $T_{\alpha\beta}$. With these semiaxes ($a$, $b$ and $c$), we can calculate the asphericity $A$ \cite{millett_effect_2009}:
\begin{eqnarray}
\label{as}
    A\equiv \frac{(a-b)^2+(b-c)^2+(c-a)^2}{2(a+b+c)^2}.
\end{eqnarray}
This quantity measures how spherical the enveloping ellipsoid is: $A=0$ indicates a perfectly spherical shape, while a rod-like ellipsoid yields $A=1$.
\subsection{\label{ssec:linkn}Linking number.}

We define two linear arms to form a single circularized arm if their ending beads are within the distance $r_b=1.2\sigma$.
Once two arms are circularized, they define, together with a line joining their anchor points, a closed curve. One could thus estimate the topological entanglement between pairs of circularized arms by computing the linking number between two such curves.

Given two  closed oriented curves in 3D $C_1$ and $C_2$, their {\it linking number} $Lk(C_1,C_2)$ is defined with the Gauss integral~\cite{tubiana_topology_2024}:
\begin{multline}
Lk(C_1,C_2) = \frac{1}{4\pi} \oint_{C_1} \oint_{C_2}
\frac{\mathbf{r}_1(s_1) - \mathbf{r}_2(s_2)}
{\lvert \mathbf{r}_1(s_1) - \mathbf{r}_2(s_2) \rvert^{3}}
\cr 
\cr
\cdot
\left[
\frac{\mathrm{d}\mathbf{r}_1(s_1)}{\mathrm{d}s_1}
\times
\frac{\mathrm{d}\mathbf{r}_2(s_2)}{\mathrm{d}s_2}
\right]
\, \mathrm{d}s_1\, \mathrm{d}s_2 .
\end{multline}
To compute the integral properly, the two curves $C_1$ and $C_2$ must not cross. However, this can happen if one connects the anchoring points of both curves in the same way, for example on the surface of the core. To avoid this issue we consider two possible schemes to join these points: a closure on the core's surface and one through its center. By construction, these cannot cross; however, for each pair of circularized arms there are two possible choices, depending on which curve is closed in a way and which in the other, resulting in two possible links. For this reason, we define the linking number $Lk$ as the average over them~\cite{breoni_conformation_2024}. To speed up the calculation, the links are further rectified through an adapted Kymoknot code~\cite{tubiana_kymoknot_2018} before computing the integral. We define the spherical angle between the two grafting points of a circularized arm and the core center as $\alpha$ (see Fig.~\ref{fig:sticky}b).

\section{Results}
\label{sec:results}
\subsection{\label{ssec:gyration}
Circularized stars' conformations reveal a discontinuous transition in bad solvent}
\begin{figure}[h]
\centering
\includegraphics[width=0.44\textwidth]{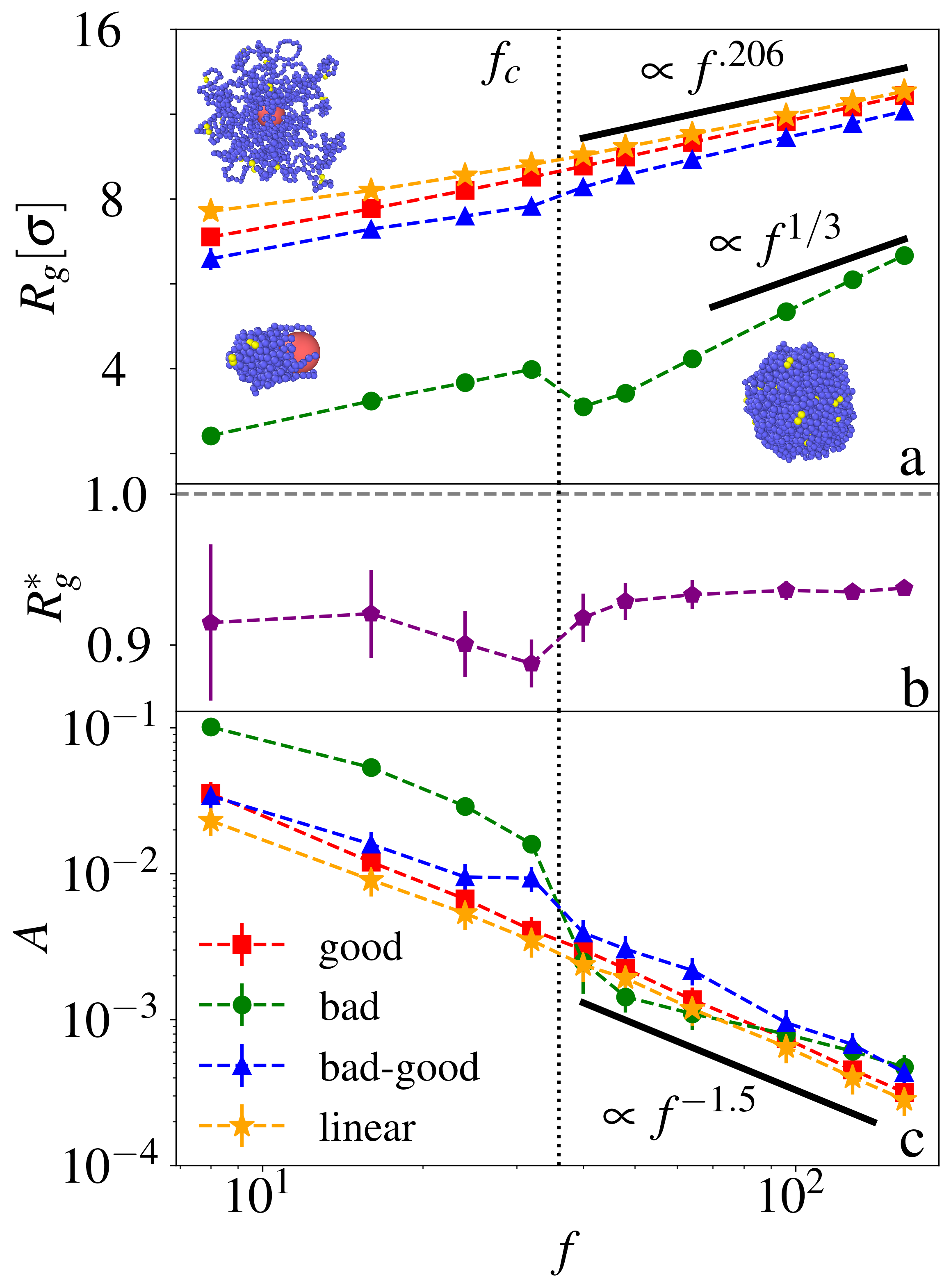}
\caption{\db{(a) Gyration radius $R_g$ as a function of $f$. The different symbols represent star polymers in different setups: linear, i.e.,  not circularized (dark yellow stars), circularized in good solvent (red squares), bad solvent (green circles) and finally circularized in bad solvent and then simulated in good solvent (blue triangles). The insets show simulation snapshots of star polymers in various conditions: good solvent with $f=32$ (upper left), bad solvent with $f=8$ (lower left) and bad solvent with $f=64$ (lower right). The vertical dotted line represents the value of $f$ at which the stars in bad solvent show a discontinuity, $f_c$. The solid black lines indicate the  $\propto f^{.206}$ regime of $R_g$ for the stars in good solvent and the $\propto f^{1/3}$ one for those in bad solvent. (b) Ratio $R_g^*$ between the $R_g$ of {\it bad-good} stars and {\it good} ones as a function of $f$. The dashed line at $R_g^*=1$ marks the good solvent case, serving as a reference. (c) Asphericity as a function of $f$. The solid black line highlights the $\propto f^{-1.5}$ behavior typical of star polymers in good solvent. All quantities are measured for star polymers with $n=40$ and $\sigma_c=6\sigma$.}}
\label{fig:rg}
\end{figure}

To understand the effect of circularization, we compare the scaling behavior for the gyration radius of circularized star polymers in various settings with that of linear star polymers (i.e. not circularized -- denoted in the following as {\it linear}) in good solvent. We reiterate that the label {\it good} refers to stars circularized in good solvent, {\it bad} to stars circularized in bad solvent and {\it bad-to-good} to stars circularized in bad solvent, quenched, and then simulated in good solvent. We can see in Fig.~\ref{fig:rg}a that, for $f$ large enough, all stars in good solvent, that is, {\it linear}, {\it good} and {\it bad-to-good}, follow the expected scaling
\begin{equation}
    R_g (\nu) \propto  n^{\nu}f^{(1-\nu)/2}= n^{.588}f^{.206},
\end{equation}
typical of stars in a {\it swollen} regime~\cite{daoud_star_1982} (with $f^{1/2}\leq n$), where $\nu=0.588$ is the Flory exponent in good solvent. Thus, the circularization affects only the scaling prefactor: stars circularized in good solvent show a smaller $R_g$ than linear stars, while {\it bad-to-good} stars display an even smaller value of $R_g$. \db{More specifically, the $R_g$ of {\it bad-to-good} stars is 7\% to 12\%  smaller than that of {\it good} stars, as shown in Fig.~\ref{fig:rg}b, where  we present the ratio $R_g^*\equiv R_g(bad-to-good)/R_g(good)$.} Polymers in bad solvent show a different scaling behavior; interestingly, $R_g$ displays a non-monotonicity as a function of $f$, with a drop appearing after a critical value $f\simeq36\equiv f_c$. This drop highlights the separation between two very different scaling regimes, neither of which agrees with the theoretical scaling of a star in bad solvent, that reads
\begin{equation}
    R_g (\nu_c) \propto n^{\nu_c}f^{(1-\nu_c)/2}= n^{1/3}f^{1/3},
\end{equation}
where $\nu_c=1/3$ is the globular exponent. Direct observation of simulation snapshots highlights the origin of this drop: for $f<f_c$ the arms of the star tend to accumulate asymmetrically on one side of the core (Fig.~\ref{fig:rg}a, lower-left inset), while for $f>f_c$ the core is uniformly covered by the arms (Fig.~\ref{fig:rg}a, lower-right inset). \db{A similar phenomenology was also observed in linear stars, where the full coverage of the core surface depends on the grafting density $\rho_g$~\cite{verso_effect_2012,selli_impact_2019}, that is, the number of arms per unit of core surface}. \db{The effects of this transition on {\it bad-to-good} polymers are highlighted in Fig.~\ref{fig:rg}b, where $R_g^*$ has a clear minimum before $f_c$.} The transition becomes even more apparent in the asphericity $A$, which can better differentiate between a macromolecule that is mostly spherical and one that is very asymmetrical: in Fig.~\ref{fig:rg}c we indeed see that the asphericity of stars in bad solvent and bad-to-good solvent deviates from the typical $\propto f^{-1.5}$ behavior\cite{breoni_conformation_2024} and shows a significant drop around $f_c$. We further note that the transition is also visible in the radial distribution function (see SI).

\subsection{\label{ssec:rlink}Topological analysis of the linked arms}

\begin{figure}[b]
\centering
\includegraphics[width=0.44\textwidth]{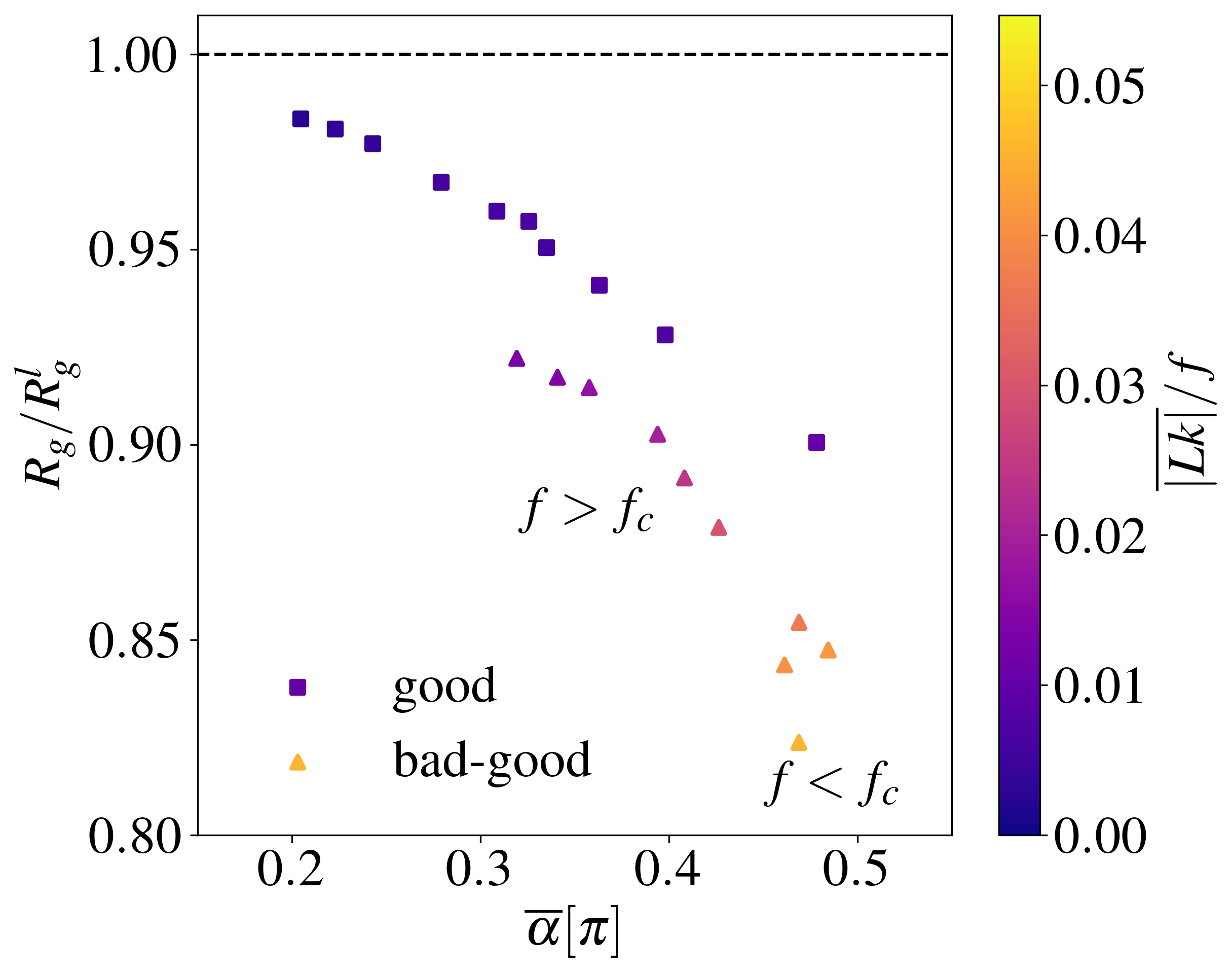}
\caption{Ratio $R_g/R_g^l$, where $R_g^l$ is the gyration radius of the linear star with the same values of $n$ and $f$, as a function of the average separation angle between circularized arms $\overline{\alpha}$ for {\it good} (squares) and {\it bad-to-good} (triangles) star polymers. The color gradient indicates the average linking number per arm $\overline{|Lk|}$ divided by $f$.}
\label{fig:link}
\end{figure}

\begin{figure*}[!t]
\centering
\includegraphics[width=0.88\textwidth]{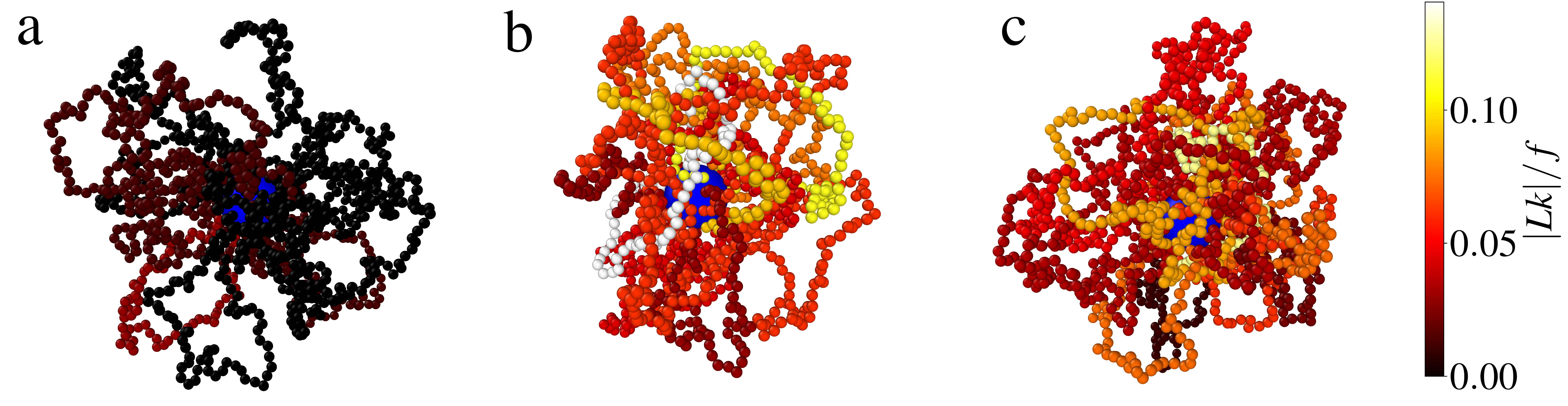}
\caption{Simulation snapshots of star polymers with core diameter $\sigma_c=6\sigma$ and number of beads per arm $n=40$ in different setups; the color gradient indicates the relative linking for each arm $|Lk|/f$, while the core is colored in blue. In the first two panels we see star polymers with number of arms $f=32<f_c$ (a) in {\it good} and (b) {\it bad-to-good} setups. (c) Star with $f=40>f_c$ in {\it bad-to-good} conditions. The snapshots highlight that the amount of relative linking is much smaller if circularization happens in good solvent with respect to bad solvent. If circularization happens in bad solvent, the $|Lk|/f$ is heavily influenced by $f$: for $f<f_c$ (b) we observe significantly more relative linking than for $f>f_c$ (c).}
\label{fig:lnk_snap}
\end{figure*}

In order to understand why {\it bad-to-good} stars show a significantly smaller $R_g$ than {\it good} star and how the transition in bad solvent further affects the gyration radius of {\it bad-to-good} stars, we study the degree of linking among arms. In fact, as shown in a previous work~\cite{breoni_conformation_2024}, linked stars present the same $R_g$ scaling of linear stars (up to a certain degree of linking) but a smaller prefactor; indeed, the extent of the reduction in $R_g$ is governed by the degree of interlinking among the arms (quantified by the average linking number $\overline{|Lk|}$) and by the average spherical angle between the grafting points of the circularized arms (the opening angle $\overline{\alpha}$). Here, $\overline{(\cdot)}$ denotes the average over all circularized arms. We observe in Fig.~\ref{fig:link} that circularization in bad solvent leads to significantly larger $\overline{|Lk|}$ and $\overline{\alpha}$, and hence to a smaller $R_g$.  We can also notice that stars formed in bad solvent show two distinct behaviors with increasing $f$: at small values of $f$ {neither the steric hindrance, due to short range monomer-monomer repulsion, nor the connectivity of the arms prevent the agglomeration of all the monomers in a single globule. Indeed, the system maximizes the energetic gain by forming this globule on one side of the central colloid, rather than around it, as the surface-to-volume ratio is highly favorable in the former case.\\ 
The formation of this globule} leads to a high degree of linking between the arms; increasing $f$ beyond $f_c$ the globule disappears and we observe, on average, a decrease in the degree of linking and smaller opening angles. 
This phenomenology can be recasted as a transition between an asymmetrically wetted and a fully wetted core in bad solvent. If all arms are able to concentrate on one side of the core they can, in principle, interact with all other arms, even those grafted on the opposite side. When this is prevented by connectivity and steric hindrance, the system minimizes its energy by uniformly wetting the core and arms tend to only interact with those in their proximity, as evidenced by the low values of $\overline{\alpha}$. Notice further that the effects of linking are maximized at values of $f$ slightly smaller than the transition functionality $f_c$, where the stars have the largest number of arms (and hence largest possible linking complexity) and each arm can bond with any one of the other arms. Indeed, in Fig.~\ref{fig:rg}b the largest relative gap between the $R_g$ of {\it good} and {\it bad-to-good} stars can be found for $f=32$, right before $f_c$. 
In Fig.~\ref{fig:lnk_snap} we showcase the relative linking $|Lk|/f$ for each arm resulting from different preparations: {\it good} stars (a) always present a significantly smaller $|Lk|/f$ with respect to their {\it bad-to-good} counterpart (b,c). Furthermore, {\it bad-to-good} circularization with $f<f_c$ (b) yields a larger $|Lk|/f$ than circularization with $f>f_c$ (c).
\subsection{\label{ssec:rasph}Characterization of the wetting transition}
\begin{figure}[h]
\centering
\includegraphics[width=0.5\textwidth]{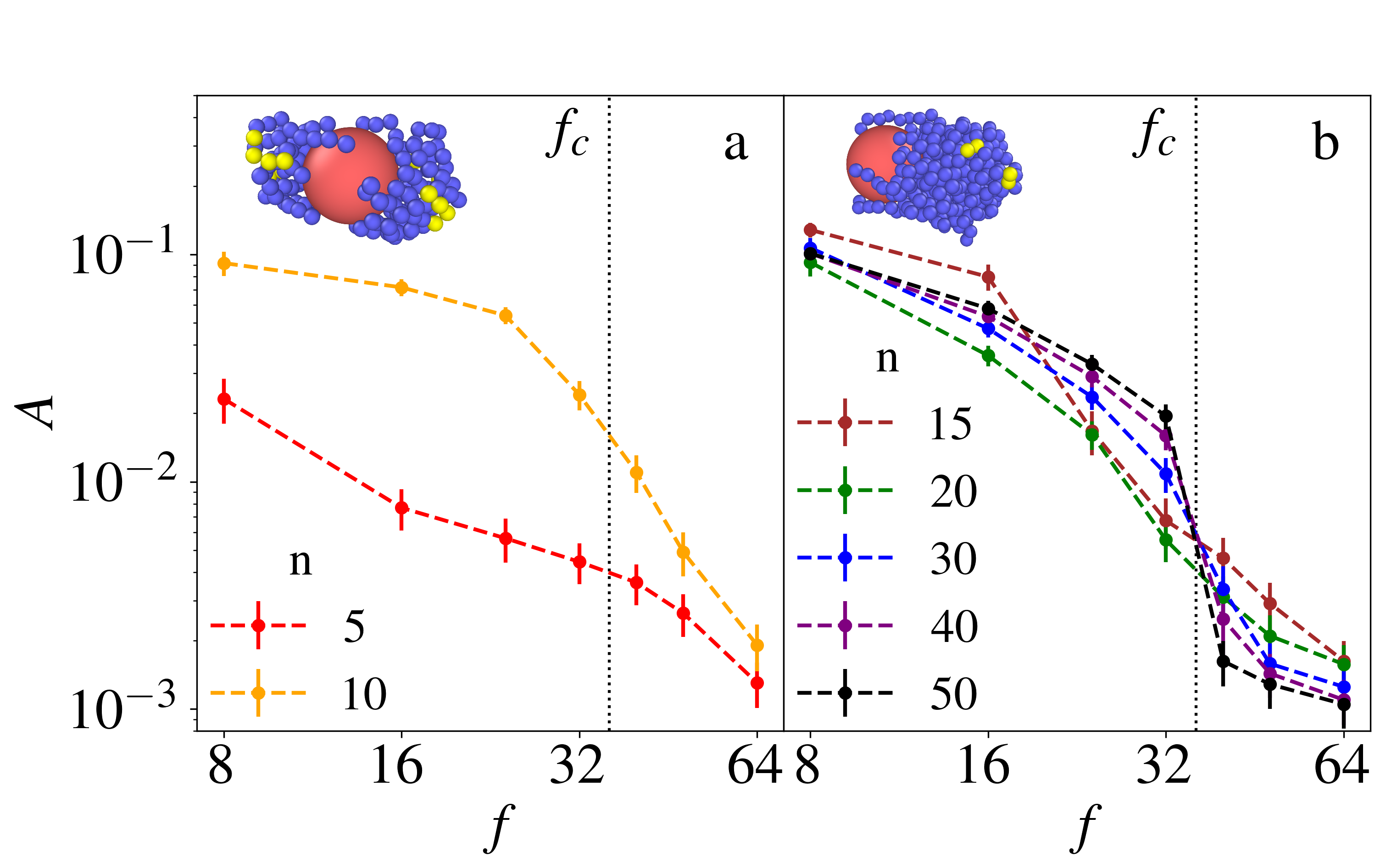}
\caption{Asphericity $A$ of star polymers in bad solvent with $\sigma_c/\sigma\equiv c=6$ as a function of $f$ for different values of $n$. In (a) we show arms with $n< n_c$, while in (b) we show arms with $n\geq n_c$. \db{The insets show simulation snapshots for the cases $n=10$, $f=16$ (a) and $n=40$, $f=8$ (b).} }
\label{fig:asph_transition}
\end{figure}

Having understood the importance that the core wetting transition has on the linking number and gyration radius of star polymers, we are interested in understanding the role played by the arms' length and by the central core. In order to determine the effect of arm length on this transition, we look at the point at which the asphericity $A$ of stars in bad solvent drops as a function of $f$, while varying $n$ in the range $5 \leq n \leq 50$ beads. Intuitively, an asymmetric collapse can take place if the arms are at least half of the core's circumference long, $n\ge n_c(c)\equiv\pi (c+1)/2$, where the addition of $\pi/2$ is needed to take into account the distance between the core and the first bead of the arms. In the case of $c=6$ we find that $n_c\gtrsim 10$: indeed for $n\leq 10$ we see that the transition is not easily identifiable (Fig.~\ref{fig:asph_transition}a). As arms cannot reach the other side of the core, the formation of a single asymmetrical clump is impossible, and instead two or more clumps form on different sides (Fig.~\ref{fig:asph_transition}a, inset).  For $n>10$ we find instead that, as the length of the arms increases, the transition becomes sharper and sharper, allowing a clear estimation of $f_c$  (Fig.~\ref{fig:asph_transition}b).

\begin{figure}[h]
\centering
\includegraphics[width=0.5\textwidth]{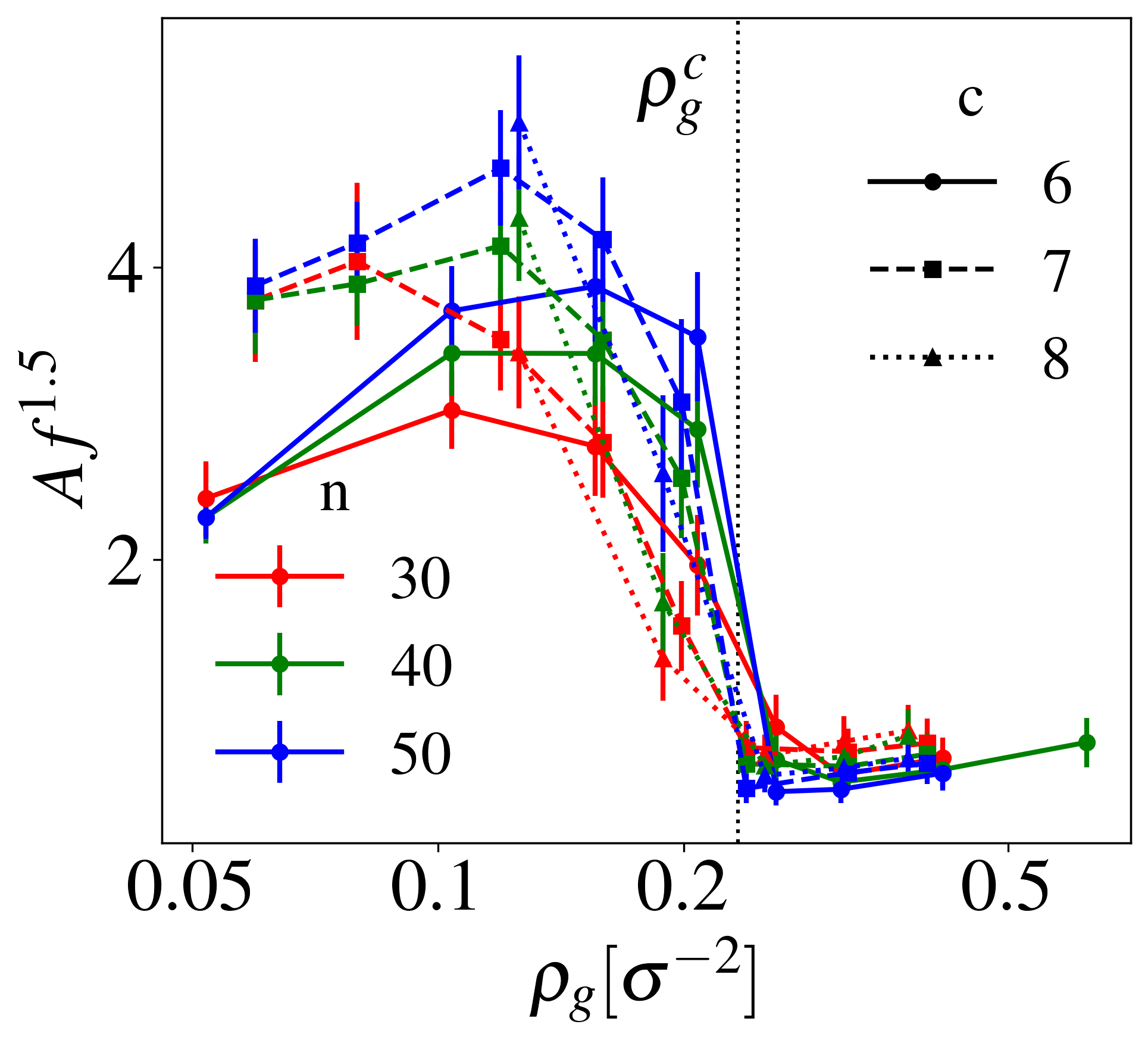}
\caption{Asphericity $A$ normalized by $f^{-1.5}$ of star polymers in bad solvent with various core ratios $c$ and arm lengths $n$ as a function of \db{the grafting density $\rho_g$. The vertical line $\rho_g^c$ denotes the critical grafting density.}}
\label{fig:asph_bad}
\end{figure}

Finally, we find that $f_c$ depends on the core size, and more specifically on the \db{grafting density $\rho_g$. Considering a ratio $\sigma_c/\sigma\equiv c$ we have that this grafting density $\rho_g(c,f)$ is equal to
\begin{equation}
    \rho_g(c,f)=\frac{f}{\pi(\sigma_c+\sigma)^2}=\frac{f}{\pi\sigma^2(c+1)^2}.
\end{equation}
As $c$ grows, we observe that the critical grafting density remains constant at $\rho_g(c,f_c)\equiv \rho_g^c\simeq 0.23\sigma^{-2}$, allowing us to conclude that
\begin{equation}
    f_c(c)\simeq 0.73(c+1)^2.
\end{equation}
In Fig.~\ref{fig:asph_bad}, we show that by plotting the asphericity as a function of the grafting density $\rho_g$, the transition points for different values of $c$ collapse together.}

\section{\label{sec:outro}Discussion and conclusions.}
We developed a model for the formation of cyclical stars with click chemistry, wherein reactive arm endings of linear stars are allowed to bond with each other. By employing a three-body potential to model these bonds, we are able to dynamically exchange them, allowing for the exploration of multiple topologies within the same run. We use such model to study the conformation and topological properties of the star polymers both in bad and good solvent and characterize the properties of cyclical stars in good solvent which were previously circularized in bad solvent. We find that the condition of the solvent in which stars are circularized has an important effect on the magnitude of linking between arms, and as a consequence their gyration radius. Indeed, worse solvent always increases linking between arms, leading in good solvent to a significantly smaller gyration radius for stars circularized in bad solvent. Furthermore, we find that stars in bad solvent undergo a wetting transition as a function of their functionalization: for $f<f_c$ arms tend to collapse on one side of the core, while for $f>f_c$ the core becomes fully wetted by the arms. Consequently, stars with $f<f_c$ present a larger linking number, as all arms can interact with each other, regardless of their anchor point on the core, unlike in the case $f>f_c$. This transition value $f_c$ is found to be independent of the length of the arms, with the condition that the length of the arm must be at least half of the core's circumference. Finally, we find that \db{the grafting density at which the transition happens is constant $\rho_g^c\simeq 0.23\sigma^{-2}$ and, as a consequence, the critical functionality goes as $f_c(c)\propto(c+1)^2$}, where $c\equiv\sigma_c/\sigma$ is the ratio of the core diameter with the diameter of an arm bead.\\
These results further enlighten the physics of cyclical star polymers and its interplay with topology, expanding on those we obtained in Ref.~\cite{breoni_conformation_2024}. Crucially, we present here a realistic path to circularization with click chemistry, which could be reproduced in experiments, and define the solvent conditions and range of values of $n$ and $f$ for which circularization can maximize the effects of linking. The effects of solvent quality on linking are moreover significant for the physics of planar brushes, where the circularization of the brushes enhances their lubricant properties\cite{romio_topological_2020}, and a significant amount of linking among them could further improve them.\\
In the future, it will be interesting to investigate if the introduction of a different species of reactive ends could open up a design space for high yield macromolecules with a specific topology, to be realized in different solvent conditions: from a numerical perspective, advanced sampling techniques can help shedding light on the interplay between bad solvent collapse and circularization~\cite{leitold2014folding} and, in general, on the folding pathway of these macromolecules~\cite{leitold2015string}.
Finally, from a more applicative perspective, star polymers have interesting applications as absorbers of pollutants\cite{roma2021theoretical,vorsmann2024colloidal}. It would be interesting to study how circularization and linking affect their absorption capability: by exploiting the phenomenology discussed in this paper, we could affect the shape and organization of these polymeric ``sponges'', possibly leading to a larger carrying capacity or improved selectivity in realistic settings, e.g., including explicit ions~\cite{tagliabue2025role}.

\begin{acknowledgments}

This work has been supported by the project “SCOPE—Selective Capture Of Metals by Polymeric spongEs” funded by the MIUR Progetti di Ricerca di Rilevante Interesse Nazionale (PRIN) Bando 2022 (Grant No. 2022RYP9YT). The authors acknowledge the CINECA award under the ISCRA initiative, for the availability of high-performance computing resources and support. D.B. and L.T. thank the UniTN HPC cluster for offering the computing resources. L.T. acknowledges support from ICSC – Centro Nazionale di Ricerca in HPC, Big Data and Quantum computing, funded by the European Union under NextGenerationEU.
\end{acknowledgments}
\section*{Author Declarations}
\subsection*{Conflict of Interest}
The authors have no conflicts to disclose.
\subsection*{Author Contributions}
\textbf{Davide Breoni:} Conceptualization (supporting); Formal Analysis (lead); Investigation (lead); Methodology (equal); Software (lead); Writing -- original draft (equal); Writing -- review \& editing (equal). \textbf{Emanuele Locatelli:} Conceptualization (equal); Formal Analysis (supporting); Funding Acquisition (equal); Investigation (supporting); Methodology (equal); Writing -- original draft (equal); Writing -- review \& editing (equal). \textbf{Luca Tubiana:} Conceptualization (equal); Formal Analysis (supporting); Funding Acquisition (equal); Investigation (supporting); Methodology (equal); Writing -- review \& editing (equal).
\section*{Data Availability Statement}
The data that support the findings of this study are available from the corresponding author upon reasonable request.
\bibliography{click,topology}
\newpage
\title{}
\clearpage
\twocolumngrid 
\appendix
\onecolumngrid 

\renewcommand{\thefigure}{S\arabic{figure}}
\setcounter{figure}{0}
\begin{center}
\textbf{\Large  \vspace*{1.5mm} The effects of solvent quality and core wetting on the circularization of star polymers -- Supplemental Information } \\
\vspace*{5mm}
Davide Breoni, Emanuele Locatelli, and Luca Tubiana
\vspace*{10mm}
\end{center}

\section{The radial density}
\begin{figure}[h]
\centering
\includegraphics[width=0.8\textwidth]{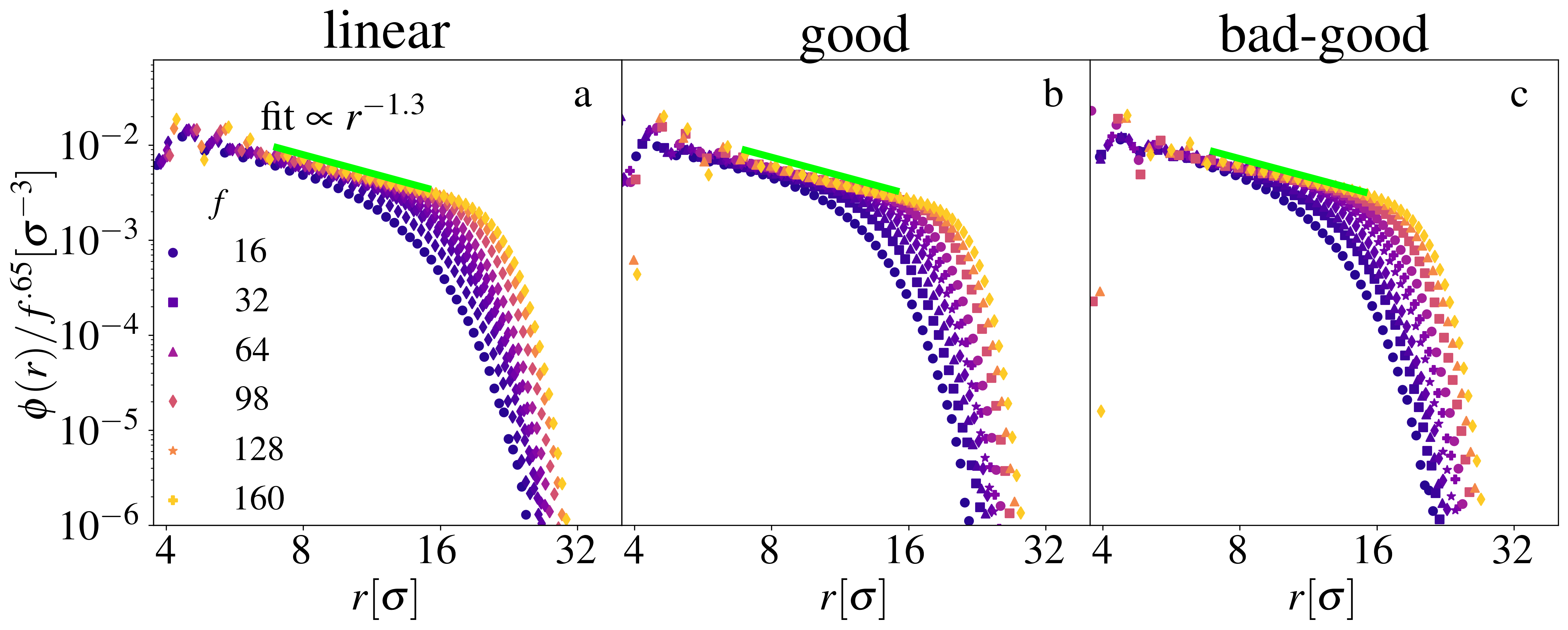}
\caption{Normalized radial density $\phi(r)/f^{.65}$ of star polymers in good solvent solvent with arm length $n=40$, core radius $\sigma_c$ for various values of $f$ in different setups: linear stars (a), stars circularized in good solvent (b) and stars circularized bad solvent and simulated in good solvent (c). }
\label{fig:rad}
\end{figure}

\begin{figure}[h]
\centering
\includegraphics[width=0.8\textwidth]{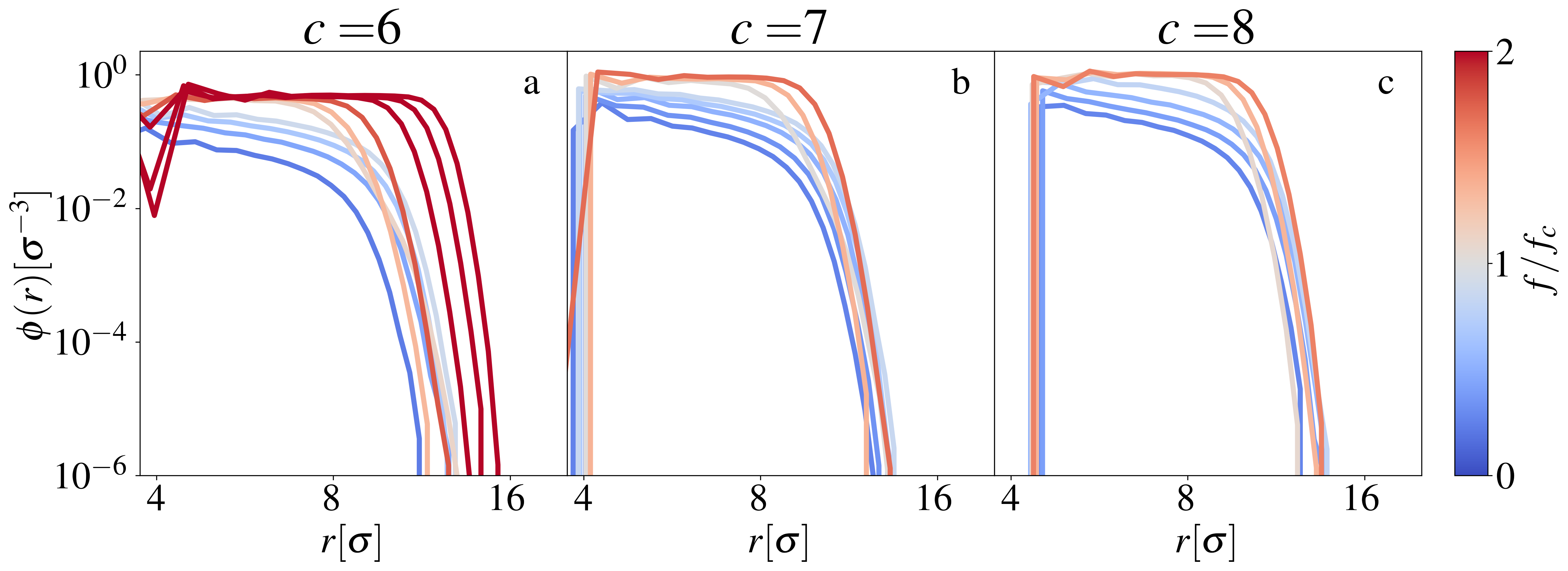}
\caption{Radial density $\phi(r)$ of star polymers in bad solvent solvent with arm length $n=40$, core radius ratio $\sigma_c/\sigma=c=6,7,8$ (a,b,c respectively) for various values of $f$.}
\label{fig:radbad}
\end{figure}

In this section, we show how the transition can be revealed by the radial density function and, in turn, what the signature of the transition on the radial density function is. We define the {\it radial density} $\phi(r)$ as:
\begin{equation}
\label{phi_r}
    \phi(r) \equiv \frac{1}{4\pi r^2}\left\langle\sum_{i=1}^{N}\delta\left( r-|\textbf{r}_i-\textbf{r}_0|\right)\right\rangle,
\end{equation}
where $N=f\cdot n $ is the total number of arm beads, $\textbf{r}_0$ is the position of the core and $\langle\cdot\rangle$ is the ensemble average, computed over time and over 8 independent realizations (see Sec.~\ref{sec:methods}).
For cyclical stars in good solvent, one can see in Fig.~\ref{fig:rad}b,c) that the scaling, both in $r$ and $f$, is always consistent with that of a linear star in the swollen regime~\cite{grest_structure_1987} (see also Fig.~\ref{fig:rad}a):
\begin{equation}
\label{eq:radprof}
    \phi(r,\nu) \propto f^{(3\nu-1)/2\nu}r^{(1-3\nu)/\nu}=f^{.650}r^{-.130}.
\end{equation}
This further confirms the results reported in Fig.~2 of the main text.
Instead, as visible in Fig.~\ref{fig:radbad}, the radial density of stars in bad solvent displays drastically different trends, both with $r$ and $f$. More specifically, we observe that there are two very different regimes, separated by $f_c$. For $f>f_c$, $\phi(r)$ remains initially constant, as expected in the globular regime ($\nu_c=1/3$):
\begin{equation}
\label{eq:radbad}
    \phi(r,\nu_c) \propto f^{(3\nu_c-1)/2\nu_c}r^{(1-3\nu_c)/\nu_c}=1,
\end{equation}
while, for $f<f_c$, $\phi(r)$ starts decaying immediately with $r$, as a consequence of the asymmetrical configurations observed before the wetting transition. 
\end{document}